\documentclass[conference]{IEEEtran}
\usepackage{cite}
\usepackage[dvips]{graphicx}
\usepackage[cmex10]{amsmath}
\usepackage{amssymb}
\usepackage{flushend}

\newcommand{\Prb}{\mathsf{P}}\newcommand{\Exp}{\mathsf{E}}
\newcommand{\C}{\mathbb{C}}\newcommand{\R}{\mathbb{R}}
\newcommand{\N}{\mathbb{N}}
\newcommand{\dd}{\mathrm{d}}\newcommand{\ee}{\mathrm{e}}
\newcommand{\B}{\mathcal{B}}
\newcommand{\Lpl}{\mathcal{L}}
\newcommand{\SIR}{\mathsf{SIR}}
\newcommand{\Gam}{\mathrm{Gam}}
\newcommand{\ind}[1]{\boldsymbol{1}_{#1}}
\newcommand{\cmb}[2]{\mathchoice
  {\genfrac{}{}{0pt}{0}{#1}{#2}}%
  {\genfrac{}{}{0pt}{1}{#1}{#2}}%
  {\genfrac{}{}{0pt}{2}{#1}{#2}}%
  {\genfrac{}{}{0pt}{3}{#1}{#2}}}
\let\Bar\overline

\newtheorem{proposition}{Proposition}
\newtheorem{theorem}{Theorem}

\newtheorem{lemma}{Lemma}
\newtheorem{example}{Example}
\newtheorem{remark}{Remark}

\begin{document}\sloppy\allowbreak\allowdisplaybreaks

\title{Downlink Coverage Probability in a Cellular Network with
  Ginibre Deployed Base Stations\\ and Nakagami-$m$ Fading Channels}

\author{%
  \IEEEauthorblockN{Naoto Miyoshi}
  \IEEEauthorblockA{Department of Mathematical and Computing Sciences\\
    Tokyo Institute of Technology\\
    2-12-1-W8-52 Ookayama, Tokyo 152-8552, Japan\\
    Email: miyoshi@is.titech.ac.jp}
  \and
 \IEEEauthorblockN{Tomoyuki Shirai}
 \IEEEauthorblockA{Institute of Mathematics for Industry\\
   Kyushu University\\
   744 Motooka, Fukuoka 819-0395, Japan\\
   Email: shirai@imi.kyushu-u.ac.jp}%
}

\maketitle

\begin{abstract}
Recently, spatial stochastic models based on determinantal point
processes~(DPP) are studied as promising models for analysis of
cellular wireless networks.
Indeed, the DPPs can express the repulsive nature of the macro base
station~(BS) configuration observed in a real cellular network and
have many desirable mathematical properties to analyze the network
performance.
However, almost all the prior works on the DPP based models assume the
Rayleigh fading while the spatial models based on Poisson point
processes have been developed to allow arbitrary distributions of
fading/shadowing propagation effects.
In order for the DPP based model to be more promising, it is essential
to extend it to allow non-Rayleigh propagation effects.
In the present paper, we propose the downlink cellular network model
where the BSs are deployed according to the Ginibre point
process, which is one of the main examples of the DPPs, over
Nakagami-$m$ fading.
For the proposed model, we derive a numerically computable form of the
coverage probability and reveal some properties of it numerically and
theoretically.
\end{abstract}

\IEEEpeerreviewmaketitle

\section{Introduction}

Spatial stochastic models for wireless communication networks have
attracted much attention~(see, e.g.,
\cite{BaccBlas09a, BaccBlas09b, Haen13, Mukh14}).
This is because the performance of a wireless network depends on the
spatial configuration of wireless nodes and spatial point processes
can capture the irregularity of the node configuration in a real
network.
In the analysis of such spatial models for wireless networks, the
stationary Poisson point processes~(PPPs) have widely used to model
the configuration of wireless nodes.
While such a PPP based model is tractable to analyze, it is an
idealized one and ignores the spatial correlation among the wireless
nodes.

Recently, spatial models based on determinantal point processes~(DPPs)
are studied as promising models for analysis of cellular wireless
networks (see, e.g., \cite{MiyoShir14a, TorrLeon14, NakaMiyo14,
  MiyoShir14b, KobaMiyo14, LiBaccDhilAndr14}).
Indeed, the DPPs can express the repulsive nature among the macro base
stations~(BSs) observed in a real cellular network and have many
desirable mathematical properties to analyze the network performance
(see, e.g., \cite{Sosh00, ShirTaka03, HougKrisPereVira09}).
However, almost all the prior works on the DPP based models assume the
Rayleigh fading while the PPP based models have been developed to
allow arbitrary distributions of fading/shadowing propagation
effects~(see, e.g., \cite{Mukh11, MadhRestLiuBrow12, KeelBlasKarr13,
  DiReGuidCora13}).
There are only a few exceptions as follows:
Although \cite{MiyoShir14a} mainly assume the Rayleigh fading, it
states in Remark~2 that the fading from the interfering BSs can be
generalized to follow an arbitrary distribution while retaining the
Rayleigh fading from the serving BS.
Also, \cite{TorrLeon14} investigates the tail asymptotics of the
distribution of interference using different fading distributions,
though it does not take into account the fading from the serving BS.

In order for the DPP based model to be more promising, it is essential
to extend it to allow non-Rayleigh propagation effects from both the
serving and interfering BSs.
In the present paper, we propose the downlink cellular network model
where the BSs are deployed according to the Ginibre point
process~(GPP) over Nakagami-$m$ fading~(see \cite{Naka60}).
The GPP is one of the main examples of the DPPs and is suited to model
the macro BS configuration~(see e.g., \cite{MiyoShir14a}).
The Nakagami-$m$ fading includes the Rayleigh one as a special case of
$m=1$ and makes the fading effects to follow the $m$th Erlang
distribution with unit mean.
It is further important to study the Erlang distributed propagation
effects since it could lead to an extension to the analysis of
multi-input multi-output heterogeneous cellular networks as in
\cite{HuanPapaVenk12, DhilKounAndr13, GuptDhilVishAndr14}.
For the proposed model, we derive a numerically computable form of
the coverage probability and reveal some properties of it numerically
and theoretically.

The paper is organized as follows:
In the next section, the network model which we consider is defined,
where the impact of the variability of fading effects on the
signal-to-interference ratio~(SIR) is discussed using some stochastic
order.
A brief description of the GPP is also provided.
In Section~\ref{sec:coverage}, we first give the basic formula of the
downlink coverage probability when the BSs are deployed according to a
general stationary point process and then we derive a numerically
computable form of it when the BSs are deployed according to the GPP.
In Section~\ref{sec:asympt}, the asymptotic properties of the downlink
coverage probability in two extreme cases $m=1$ and $m\to\infty$ are
discussed.
The results of some numerical experiments are shown in
Section~\ref{sec:experiments}.

\section{Network model}

\subsection{Macro base station network}

The downlink cellular network model which we consider mainly
follows the ones in~\cite{AndrBaccGant11, MiyoShir14a} and consists of
homogeneous macro BSs.
Let $\Phi$ denote a point process on $\R^2$ and let $X_i$, $i\in\N$,
denote the points of $\Phi$, where the order of $X_1$, $X_2$, \ldots,
is arbitrary.
The point process $\Phi$ expresses the configuration of the BSs and we
refer to the BS located at $X_i$ as BS~$i$.
We assume that $\Phi$ is simple and locally finite a.s.\ and also
stationary.
Each mobile user is associated with the closest BS.
Thus, due to the stationarity of the point process and the homogeneity
of the BSs, we can focus on a typical user located at the
origin~$o=(0,0)$.
For the simplicity, we limit ourselves to the interference limited
(noise-free) case throughout the paper.
It is known that, in the interference limited case of the stationary
single tier network model, the downlink coverage probability does not
depend on the intensity of the point process, transmission power and
path-loss coefficient (see, e.g., \cite{MiyoShir14a, AndrBaccGant11}),
so that, we set these values as $\pi^{-1}$, $1$ and $1$, respectively,
for convenience.
The path-loss function $\ell$ representing the attenuation of signals
with distance is then given by $\ell(r)=r^{-2\,\beta}$, $r>0$, for
some $\beta>1$.
We assume the Nakagami-$m$ fading, so that the random propagation
effect from the BS~$i$ to the typical user, denoted by $H_i$, follows
the $m$th Erlang distribution with unit mean, denoted by
$H_i\sim\Gam(m,1/m)$, $i\in\N$, where $H_i$, $i\in\N$, are mutually
independent and also independent of the point process~$\Phi$.
The shadowing is ignored in our model.

In this setting, the downlink SIR of the typical user is given by
\begin{equation}\label{eq:SIR}
  \SIR_o = \frac{H_{B_o}\,\ell(\|X_{B_o}\|)}{I_o(B_o)},
\end{equation}
where $B_o$ denotes the index of the BS associated with the typical
user at the origin; that is, $\{B_o=i\} = \{\|X_i\| \le \|X_j\|,
j\in\N\}$, and $I_o(i)$, $i\in\N$, denotes the cumulative interference
received by the typical user from all the BSs except BS~$i$; that is,
\begin{equation}\label{eq:interference}
  I_o(i) = \sum_{j\in\N\setminus\{i\}} H_j\,\ell(\|X_j\|).
\end{equation}
When we emphasize the $m$ of the Nakagami-$m$ fading, we put
the superscript~$(m)$ and write $\SIR_o^{(m)}$, $I_o^{(m)}(i)$ and so
on.

Before assigning a specific point process to the BS configuration~$\Phi$, we
here discuss the variability of the SIR in terms of the value of $m$
of the Nakagami-$m$ fading using some stochastic order.
For two random variables $X$ and $Y$ with finite expectations, we say
that $X$ is less than $Y$ in the convex order [resp.\ increasing
  convex order] and write $X\le_{\rm cx}Y$ [resp.\ $X\le_{\rm icx}Y$]
if $\Exp f(X) \le \Exp f(Y)$ for any convex [resp.\ increasing and
  convex] function~$f$ such that the expectations exist~(see, e.g.,
\cite{MullStoy02}).
It is well known that $\Gam(m,1/m)$ is decreasing in $m$ in the convex
order; that is, $m_1<m_2$ implies $H^{(m_1)} \ge_{\rm cx}H^{(m_2)}$ for
$H^{(m)}\sim\Gam(m,1/m)$.
However, we have to care about the application of the convex or increasing
convex order to the SIR in~\eqref{eq:SIR} since it may not have the
finite expectation when the path-loss function is unbounded at the
origin.
We thus consider the conditional SIR given the common BS
configuration~$\Phi$.
The following lemma implies that more variable fading effects lead to
larger and more variable conditional SIR given the common BS
configuration.

\begin{lemma}\label{lem:icx}
For each $k=1$, $2$, let $\{H_i^{(k)}\}_{i\in\N}$ denote a sequence of
mutually independent random variables representing the fading
effects.
Let also $\SIR_o^{(k)}$, $k=1$, $2$, denote the SIR of the typical
user, given by \eqref{eq:SIR}, corresponding to the fading effects
$\{H_i^{(k)}\}_{i\in\N}$ with the common BS
configuration~$\Phi=\{X_i\}_{i\in\N}$.
Then, $H_i^{(1)} \ge_{\rm cx} H_i^{(2)}$, $i\in\N$, implies
$(\SIR_o^{(1)}\mid \Phi) \ge_{\rm icx} (\SIR_o^{(2)} \mid \Phi)$ given
$\Phi$.
\end{lemma}

\begin{IEEEproof}
By Strassen's theorem~(see, e.g., \cite[Sec.~1.5]{MullStoy02}), we can
define $\{H_i^{(k)}\}_{i\in\N}$, $k=1$, $2$, satisfying $H_i^{(2)} =
\Exp(H_i^{(1)} \mid H_i^{(2)})$ a.s.\ for each $i\in\N$.
Thus, by \eqref{eq:SIR} and Jensen's inequality,
\begin{align*}
  \SIR_o^{(2)}
  &= \sum_{i\in\N}
       \frac{\Exp(H_i^{(1)} \mid H_i^{(2)})\,\ell(\|X_i\|)}
            {\sum_{j\in\N\setminus\{i\}}
               \Exp(H_j^{(1)} \mid H_j^{(2)})\,\ell(\|X_j\|)}
  \\
  &\quad\qquad\mbox{}\times
       \ind{\{\|X_i\|\le\|X_j\|,\, j\in\N\}}
  \\
  &\le \Exp\bigl(
         \SIR_o^{(1)}
       \bigm|
         \{H_i^{(2)}\}_{i\in\N}, \Phi
       \bigr)
  \quad\text{a.s.}
\end{align*}
Let $f$ denote any increasing and convex function satisfying
$\Exp\bigl(f(\SIR_o^{(1)}) \mid \Phi\bigr)<\infty$ and
$\Exp\bigl(f(\SIR_o^{(2)}) \mid \Phi \bigr)>-\infty$ a.s.
Then, Jensen's inequality implies
\begin{align*}
  \Exp\bigl(f(\SIR_o^{(2)}) \mid \Phi\bigr)
  &\le \Exp\bigl(f\bigl(
         \Exp\bigl(
           \SIR_o^{(1)}
         \bigm|
           \{H_i^{(2)}\}_{i\in\N}, \Phi
         \bigr)
       \bigr) \mid \Phi\bigr)
  \\
  &\le \Exp\bigl(f(\SIR_o^{(1)}) \mid \Phi\bigr)
   \quad\text{a.s.}
\end{align*}
\end{IEEEproof}

\subsection{The Ginibre point process}

We here give a brief description of the GPP (see, e.g., \cite{Sosh00,
  ShirTaka03, HougKrisPereVira09} for details).
The GPP is one of the DPPs on the complex plane~$\C$ defined as
follows.
Let $\Phi$ denote a simple point process on $\C$, and let $\rho_n$:
$\C^n\to\R_+$, $n\in\N$, denote its $n$th joint intensities
(correlation functions) with respect to some Radon measure $\nu$ on
$\C$; that is, for any symmetric and continuous function $f$ on $\C^n$
with compact support,
\begin{align}\label{eq:jointintensity}
  &\Exp\biggl(
     \sum_{\substack{X_1,\ldots,X_n\in\Phi\\\text{distinct}}}
       f(X_1,X_2,\ldots,X_n)
   \biggr)
  \nonumber\\
  &= \int\!\int\cdots\int_{\C^n}
       f(z_1,z_2,\ldots,z_n)
  \nonumber\\
  &\qquad\mbox{}\times
       \rho_n(z_1,z_2,\ldots,z_n)\,
     \nu(\dd z_1)\,\nu(\dd z_2)\cdots\nu(\dd z_n).
\end{align}
The point process $\Phi$ is said to be a DPP with kernel~$K$:
$\C^2\to\C$ with respect to the reference measure $\nu$ if $\rho_n$ in
\eqref{eq:jointintensity} is given by
\[
  \rho_n(z_1,z_2,\ldots,z_n) = \det(K(z_i,z_j))_{i,j=1}^n,
\]
where $\det$ denotes the determinant.
Furthermore, a DPP~$\Phi$ is said to be the GPP when the kernel is
given by $K(z,w)=\ee^{z\Bar{w}}$, $z$, $w\in\C$, with respect to the
Gaussian measure $\nu(\dd z) = \pi^{-1}\,\ee^{-|z|^2}\,\mu(\dd z)$,
where $\Bar{w}$ denotes the complex conjugate of $w\in\C$ and $\mu$
denotes the Lebesgue measure on $\C$.
It is known that the GPP is motion invariant (stationary and
isotropic) and, by the definition, its intensity is equal to
$\pi^{-1}$; that is, $\Exp\Phi(C) = \pi^{-1}\,\mu(C)$ for
$C\in\B(\C)$.
One of the useful properties of the GPP is given by the following
proposition.

\begin{proposition}[Kostlan~\cite{Kost92}]\label{prp:Kostlan}
Let $X_i$, $i\in\N$, denote the points of the GPP.
Then, the set $\{\|X_i\|^2\}_{i\in\N}$ has the same distribution as
$\{Y_i\}_{i\in\N}$, where $Y_i$, $i\in\N$, are mutually independent
and each $Y_i$ follows the $i$th Erlang distribution with unit-rate
parameter, denoted by $Y_i\sim\Gam(i,1)$, $i\in\N$.
\end{proposition}

\section{Downlink coverage probability}\label{sec:coverage}

The downlink coverage probability is defined as the probability with
which the SIR of the typical user achieves a target threshold; that
is, $\Prb(\SIR_o>\theta)$ for $\theta>0$.
In this section, we first derive the basic formula concerning the
downlink coverage probability when the BSs are deployed according to a
general stationary point process, and then we provide its numerically
computable form when the BSs are deployed according to the GPP.

\subsection{General stationary base station deployments}

\begin{lemma}\label{lem:1}
Consider the downlink cellular network model such that the BSs have
mutually independent Nakagami-$m$ fading channels and are deployed
according to a stationary point process~$\Phi$ on $\R^2$.
Then, the coverage probability for a typical user satisfies
\begin{align}\label{eq:lem1}
  &\Prb(\SIR_o^{(m)} > \theta)
  \nonumber\\
  &= \sum_{n=0}^{m-1}
       \!
       \frac{(-1)^n}{n!}\,
       \Exp\biggl[
         \frac{\dd^n}{\dd x^n}
           \!\!
           \prod_{j\in\N\setminus\{B_o\}}
           \!
             \Bigl(
               1 + \theta\,\frac{\ell(\|X_j\|)}{\ell(\|X_{B_o}\|)}\,x
            \Bigr)^{-m}
         \biggr|_{x=1}
       \biggr].
\end{align}
\end{lemma}

\begin{IEEEproof}
Throughout the proof, we fix the value of $m$ and drop off the
superscripts~$(m)$.
We have from \eqref{eq:SIR} that
\begin{align}\label{eq:lem1pr1}
  &\Prb(\SIR_o > \theta)
   = \sum_{i\in\N} \Prb(\SIR_o > \theta, \: B_o = i)
  \nonumber\\
  &= \sum_{i\in\N}
       \Prb\Bigl(
         H_i > \frac{\theta\,I_o(i)}{\ell(\|X_i\|)}, \:
         \|X_i\|\le \|X_j\|, \: j\in\N
       \Bigr).
\end{align}
For each $i\in\N$, since $H_i\sim\Gam(m,1/m)$, conditioning on
$\Phi=\{X_j\}_{j\in\N}$ and $\{H_j\}_{j\in\N\setminus\{i\}}$ yields
\begin{align}\label{eq:lem1pr2}
  &\Prb\Bigl(
     H_i > \frac{\theta\,I_o(i)}{\ell(\|X_i\|)}
   \Bigm|
     \Phi,\: \{H_j\}_{j\in\N\setminus\{i\}}
   \Bigr)
  \nonumber\\
  &= \sum_{n=0}^{m-1}
       \frac{1}{n!}\,
       \Bigl(
         \frac{m\,\theta\,I_o(i)}{\ell(\|X_i\|)}
       \Bigr)^n\,
       \exp\Bigl(
         - \frac{m\,\theta\,I_o(i)}{\ell(\|X_i\|)}
       \Bigr)
  \nonumber\\
  &= \sum_{n=0}^{m-1}
       \frac{(-1)^n}{n!}\,
       \frac{\dd^n}{\dd x^n}
         \exp\Bigl(
           - \frac{m\,\theta\,I_o(i)}{\ell(\|X_i\|)}\,x
         \Bigr)
       \biggr|_{x=1}.
\end{align}
Here, let us take the conditional expectation given $\Phi$.
Since $\exp\bigl(-m\,\theta\,I_o(i)\,x/\ell(\|X_i\|)\bigr)$ is
continuous in $x$ and each $H_j\sim\Gam(m, 1/m)$ has the finite moment
of any order, we can exchange the conditional expectation and derivative.
Thus, since $H_j$, $j\in\N$, are mutually independent, applying
\eqref{eq:interference} yields
\begin{align}\label{eq:lem1pr3}
  &\Exp\Bigl(
     \frac{\dd^n}{\dd x^n}
       \exp\Bigl(
         - \frac{m\,\theta\,I_o(i)}{\ell(\|X_i\|)}\,x
       \Bigr)
   \Bigm| \Phi \Bigr)
  \nonumber\\
  &= \frac{\dd^n}{\dd x^n}
       \prod_{j\in\N\setminus\{i\}}
         \Exp\Bigl(
           \exp\Bigl(
             - \frac{m\,\theta\,H_j\,\ell(\|X_j\|)}{\ell(\|X_i\|)}\,x
           \Bigr)
         \Bigm| \Phi \Bigr)
  \nonumber\\
  &= \frac{\dd^n}{\dd x^n}
       \prod_{j\in\N\setminus\{i\}}
         \Bigl(
           1 + \theta\,\frac{\ell(\|X_j\|)}{\ell(\|X_i\|)}\,x
         \Bigr)^{-m},
\end{align}
where the Laplace transform $\Lpl_H(s) = (1+s/m)^{-m}$ of
$H_j\sim\Gam(m,1/m)$ is applied in the second equality.
Finally, applying \eqref{eq:lem1pr2} and \eqref{eq:lem1pr3} to
\eqref{eq:lem1pr1}, we obtain \eqref{eq:lem1}.
\end{IEEEproof}

\subsection{Ginibre base station deployments}

We consider the GPP as the BS configuration~$\Phi$, where a point
$z=x+\mathrm{i}\,y\in\C$ is identified as $(x,y)\in\R^2$.

\begin{theorem}\label{thm1}
Consider the downlink cellular network model with path-loss function
$\ell(r)=r^{-2\beta}$, $r>0$, such that the BSs have mutually
independent Nakagami-$m$ fading channels and are deployed according to
the GPP.
Then, the coverage probability for a typical user is given by
\begin{align}\label{eq:thm1-1}
  &\Prb(\SIR_o^{(m)} > \theta)
  \nonumber\\
  &= \int_0^\infty
       \ee^{-u}\,M^{(m)}(u)\,
       \sum_{n=0}^{m-1}\theta^n\sum_{k=0}^n
         \biggl[
           \sum_{h=1}^k (-1)^h\,h!\,S_{k,h}^{(m)}(u)
         \biggr]
  \nonumber\\
  &\mbox{}\times\!\!
         \sum_{(h_1,h_2,\ldots,h_{n-k})\in\mathcal{P}_{n-k}}
           \prod_{r=1}^{n-k}
             \frac{1}{h_r!}\,
             \biggl[
               \sum_{q=0}^{r-1}
                 (-1)^q\,q!\,T_{r,q+1}^{(m)}(u)
             \biggr]^{h_r}
     \dd u,
\end{align}
where $\sum_{h=1}^0a_h=a_0$, 
$\mathcal{P}_k = \bigl\{(h_1,h_2,\ldots,h_k)\in (\N\cup\{0\})^k \mid
\sum_{r=1}^k r\,h_r=k \bigr\}$, and
\begin{align}
  &M^{(m)}(u)
   = \prod_{i=0}^\infty J_i^{(m,0)}(u),
  \label{eq:thm1-2}
  \\
  &S_{k,h}^{(m)}(u)
   = \sum_{i=0}^\infty
       \frac{u^i}{i!}\,\bigl(J_i^{(m,0)}(u)\bigr)^{-h-1}\,
       V_{k,h,i}^{(m)}(u),
  \label{eq:thm1-3}
  \\
  &T_{k,h}^{(m)}(u)
   = \sum_{i=0}^\infty
       \bigl(J_i^{(m,0)}(u)\bigr)^{-h}\,
       V_{k,h,i}^{(m)}(u),
  \label{eq:thm1-4}
\end{align}
with
\begin{align}\label{eq:thm1-6}
  V_{k,h,i}^{(m)}(u)
  &= \sum_{(r_1,\ldots,r_{k-h+1})\in\mathcal{Q}_{k,h}}
       \prod_{q=1}^{k-h+1}
         \frac{1}{r_q!}
  \nonumber\\
  &\qquad\qquad\mbox{}\times
         \Bigl[
           \Bigl(\cmb{m+q-1}{m-1}\Bigr)\,
           J_i^{(m,q)}(u)
         \Bigr]^{r_q},
\end{align}
$\mathcal{Q}_{k,h} =
\bigl\{(r_1,r_2,\ldots,r_{k-h+1})\in(\N\cup\{0\})^{k-h+1} \mid
\sum_{q=1}^{k-h+1}q\,r_q = k, \: \sum_{q=1}^{k-h+1} r_q = h \bigr\}$,
and
\begin{equation}\label{eq:thm1-5}
  J_i^{(m,q)}(u)
  = \frac{1}{i!}
    \int_u^\infty
      \frac{\ee^{-y}\,y^i\,(u/y)^{q\beta}}
           {\bigl(1+\theta\,(u/y)^{\beta}\bigr)^{m+q}}\,
     \dd y,
  \;\; q=0,1,2,\ldots.
\end{equation}
\end{theorem}

\begin{remark}
The numerical computation of \eqref{eq:thm1-1}--\eqref{eq:thm1-5}
seems complicated.
Note, however, that the infinite product and infinite sums in
\eqref{eq:thm1-2}--\eqref{eq:thm1-4} are not nested.
The computation is thus scalable in the sense that these infinite
product and infinite sums can be computed simultaneously in one
iteration scheme.
The case of $m=1$ in Theorem~\ref{thm1}, of course, coincides with
Theorem~1 in~\cite{MiyoShir14a}.
The cases of $m=2$ and $3$ are provided after the proof.
\end{remark}

\begin{IEEEproof}
As in the proof of Lemma~\ref{lem:1}, we drop off the
superscripts~$(m)$ and write, for example, $J_i^{(q)}$ for $J_i^{(m,
  q)}$.
By Proposition~\ref{prp:Kostlan}, we can set $\{\|X_i\|\}_{i\in\N}$
satisfying $\|X_i\|^2 \sim Y_i \sim \Gam(i,1)$ for each $i\in\N$,
where $Y_i$, $i\in\N$, are mutually independent.
Thus, applying the density functions of $\Gam(i,1)$, $i\in\N$, and
$\ell(r) = r^{-2\beta}$, $r>0$ to \eqref{eq:lem1} yields
\begin{align}\label{eq:thm1pr1}
  &\Prb(\SIR_o > \theta)
  \nonumber\\
  &= \sum_{n=0}^{m-1}
       \frac{(-1)^n}{n!}
       \sum_{i=1}^\infty
         \Exp\biggl[
           \frac{\dd^n}{\dd x^n}
             \prod_{j\in\N\setminus\{i\}}
               \Exp\Bigl[
                 \Bigl(
                   1 + \theta\,\Bigl(\frac{Y_i}{Y_j}\Bigr)^{\beta}\,x
                 \Bigr)^{-m}\,
  \nonumber\\
  &\quad\qquad\mbox{}\times
                 \ind{\{Y_j>Y_i\}}
               \Bigm| Y_i \Bigr]
           \biggr|_{x=1}
         \biggr]
  \nonumber\\
  &= \sum_{n=0}^{m-1}
       \frac{(-1)^n}{n!}
       \sum_{i=0}^\infty
         \frac{1}{i!}
         \int_0^\infty
           \ee^{-u}\,u^i\,
           \frac{\dd^n}{\dd x^n}
             \prod_{\substack{j=0\\j\ne i}}^\infty
               C_j(x; u)
           \biggr|_{x=1}\,
         \dd u,
\end{align}
where
\[
  C_j(x; u)
  = \frac{1}{j!}
    \int_u^\infty
      \frac{\ee^{-y}\,y^j}
           {\bigl(1+\theta\,(u/y)^{\beta}\,x\bigr)^m}\,
    \dd y.
\]
Note here that $C_j(1; u) = J_j^{(0)}(u)$ in \eqref{eq:thm1-5}.
The general Leibniz rule~(see, e.g., \cite[Sec.~5.2]{Olve00})
leads to
\begin{align}\label{eq:thm1pr2}
  &\frac{\dd^n}{\dd x^n}
     \prod_{\substack{j=0\\j\ne i}}^\infty C_j(x; u)
   = \frac{\dd^n}{\dd x^n}\Bigl[
       \bigl(C_i(x; u)\bigr)^{-1}
       \prod_{j=0}^\infty C_j(x; u)
     \Bigr]
  \nonumber\\
  &= \sum_{k=0}^n
       \Bigl(\cmb{n}{k}\Bigr)\,
       \frac{\dd^k}{\dd x^k}\bigl(C_i(x; u)\bigr)^{-1}\,
       \frac{\dd^{n-k}}{\dd x^{n-k}}\prod_{j=0}^\infty C_j(x; u).
\end{align}
Consider the first derivative on the right-hand side above.
It is certainly $\bigl(C_i(x; u)\bigr)^{-1}$ when $k=0$.
For $k=1,2,\ldots$, Fa\`{a} di Bruno's formula~(see, e.g.,
\cite{Roma80}) leads to
\begin{align}\label{eq:thm1pr11}
  &\frac{\dd^k}{\dd x^k}\bigl(C_i(x; u)\bigr)^{-1}
   = \sum_{h=1}^k
       (-1)^h\,h!\,\bigl(C_i(x; u)\bigr)^{-h-1}\,
  \nonumber\\
  &\mbox{}\times
       B_{k,h}\bigl(
         C_i^{(1)}(x; u),\: C_i^{(2)}(x; u),\ldots, C_i^{(k-h+1)}(x; u)
       \bigr),
\end{align}
where $B_{k,h}$ denotes the Bell polynomial;
\begin{align*}
  &B_{k,h}(x_1,x_2,\ldots,x_{k-h+1})
  \\
  &= k!
     \sum_{(r_1,\ldots,r_{k-h+1})\in\mathcal{Q}_{k,h}}
       \prod_{q=1}^{k-h+1}
         \frac{1}{r_q!}\,
         \Bigl(\frac{x_q}{q!}\Bigr)^{r_q},
\end{align*}
and for $q=1,2,\ldots$,
\begin{align*}
  C_i^{(q)}(x; u)
  &= (-1)^q\,\theta^q\,\frac{(m+q-1)!}{(m-1)!}\,
  \\
  &\quad\mbox{}\times
     \frac{1}{i!}
     \int_u^\infty
       \frac{\ee^{-y}\,y^i\,(u/y)^{q\beta}}
            {\bigl(1 + \theta\,(u/y)^{\beta}\,x\bigr)^{m+q}}\,
     \dd y.
\end{align*}
Note here that
\[
  \frac{1}{q!}C_i^{(q)}(1; u)
  = (-1)^q\,\theta^q\,
    \Bigl(\cmb{m+q-1}{m-1}\Bigr)\,J_i^{(q)}(u),
\]
so that, noting that $\sum_{q=1}^{k-h+1}q\,r_q=k$ and
applying~\eqref{eq:thm1-6}, we have
\begin{align*}
  &B_{k,h}\bigl(
     C_i^{(1)}(1; u),\: C_i^{(2)}(1; u),\ldots, C_i^{(k-h+1)}(1; u)
   \bigr)
  \\
  &= (-1)^k\,\theta^k\,k!\,V_{k,h,i}(u).
\end{align*}
Hence, applying this to \eqref{eq:thm1pr11}, multiplying $u^i/i!$ and
summing over $i=0$, $1$, \ldots (see \eqref{eq:thm1pr1}), we have
\begin{align}\label{eq:thm1pr7}
  &\sum_{i=0}^\infty
     \frac{u^i}{i!}\,
     \frac{\dd^k}{\dd x^k}\bigl(C_i(x; u)\bigr)^{-1}\biggr|_{x=1}
  \nonumber\\
  &= (-1)^k\,\theta^k\,k!
     \sum_{h=1}^k
       (-1)^h\,h!\,
       S_{k,h}(u).
\end{align}

On the other hand, we apply Fa\`{a} di Bruno's formula twice to the
second derivative on the right-hand side of \eqref{eq:thm1pr2}; first
to $\prod C_j(x; u) = \exp\bigl(\sum\log C_j(x; u)\bigr)$ and then to
$\log C_j(x; u)$.
We then have
\begin{align}\label{eq:thm1pr8}
  &\frac{\dd^{n-k}}{\dd x^{n-k}}\prod_{j=0}^\infty C_j(x; u)
   = (n-k)! \prod_{j=0}^\infty C_j(x; u)
  \nonumber\\
  &\mbox{}\times\!
       \sum_{(h_1,h_2,\ldots,h_{n-k})\in\mathcal{P}_{n-k}}
         \prod_{r=1}^{n-k}\frac{1}{h_r!}
           \biggl[
             \frac{1}{r!}
             \sum_{j=0}^\infty
               \frac{\dd^r}{\dd x^r}\log C_j(x; u)
           \biggr]^{h_r},
\end{align}
and
\begin{align}\label{eq:thm1pr9}
  &\frac{\dd^r}{\dd x^r}\log C_j(x; u)
   = \sum_{q=1}^r(-1)^{q-1}\,(q-1)!\,
       \bigl(C_j(x; u)\bigr)^{-q}\,
  \nonumber\\
  &\mbox{}\times
      B_{r,q}\bigl(
        C_j^{(1)}(x; u),\: C_j^{(2)}(x; u),\ldots, C_j^{(r-q+1)}(x; u)
      \bigr).
\end{align}
Here, taking $x=1$ reduces to
\begin{align*}
  &B_{r,q}\bigl(
     C_j^{(1)}(1; u),\: C_j^{(2)}(1; u),\ldots, C_j^{(s-r+1)}(1; u)
   \bigr)
  \\
  &= (-1)^r\,r!\,\theta^r\,V_{r,q,j}(u),
\end{align*}
so that, applying this to \eqref{eq:thm1pr9} and \eqref{eq:thm1pr8} with
$\sum_{r=1}^{n-k}r\,h_r=n-k$, we have
\begin{align}\label{eq:thm1pr10}
  &\frac{\dd^{n-k}}{\dd x^{n-k}}\prod_{j=0}^\infty C_j(x; u)
   \biggr|_{x=1}
   = (-1)^{n-k}\,\theta^{n-k}\,(n-k)!\,M(u)
  \nonumber\\
  &\mbox{}\times
      \sum_{(h_1,h_2,\ldots,h_{n-k})\in\mathcal{P}_{n-k}}
      \prod_{r=1}^{n-k}
        \frac{1}{h_r!}\,
        \biggl[
          \sum_{q=0}^{r-1}
            (-1)^q\,q!\,T_{r,q+1}(u)
        \biggr]^{h_r}.
\end{align}
Finally, we obtain \eqref{eq:thm1-1} from \eqref{eq:thm1pr1},
\eqref{eq:thm1pr2}, \eqref{eq:thm1pr7} and \eqref{eq:thm1pr10}.
\end{IEEEproof}

\begin{example}
We here give the examples of Theorem~\ref{thm1} for $m=2$ and $3$.
When $m=2$, it reduces to
\begin{align*}
  &\Prb(\SIR_o^{(2)} > \theta)
   = \int_0^\infty
       \ee^{-u}\,
       M^{(2)}(u)\,
  \nonumber\\
  &\qquad\mbox{}\times
       \bigl(
         S_{0,0}^{(2)}(u)
         + \theta\,(S_{0,0}^{(2)}(u)\,T_{1,1}^{(2)}(u) - S_{1,1}^{(2)}(u))
       \bigr)\,
     \dd u,
\end{align*}
where
\begin{align*}
  S_{0,0}^{(2)}(u)
  &= \sum_{i=0}^\infty
       \frac{u^i}{i!\,J_i^{(2,0)}(u)},
  \\
  S_{1,1}^{(2)}(u)
  &= 2\sum_{i=0}^\infty
       \frac{u^i\,J_i^{(2,1)}(u)}
            {i!\,\bigl(J_i^{(2,0)}(u)\bigr)^2},
  \\
  T_{1,1}^{(2)}(u)
  &= 2\sum_{i=0}^\infty
       \frac{J_i^{(2,1)}(u)}
            {J_i^{(2,0)}(u)}.
\end{align*}
Also, when $m=3$, it reduces to
\begin{align*}
  &\Prb(\SIR_o^{(3)} > \theta)
   = \int_0^\infty
       \ee^{-u}\,
       M^{(3)}(u)\,
  \\
  &\quad\mbox{}\times
       \biggl[
         S_{0,0}^{(3)}(u)
         + \theta\,\Bigl(
             S_{0,0}^{(3)}(u)\,T_{1,1}^{(3)}(u) - S_{1,1}^{(3)}(u)
           \Bigr)
  \\
  &\qquad\mbox{}
         + \theta^2\,\biggl(
             S_{0,0}^{(3)}(u)\,\biggl(
               \frac{\bigl(T_{1,1}^{(3)}(u)\bigr)^2}{2}
               + T_{2,1}^{(3)}(u) - T_{2,2}^{(3)}(u)
             \biggr)
  \\
  &\quad\qquad\qquad\mbox{}
             -S_{1,1}^{(3)}(u)\,T_{1,1}^{(3)}(u)
             - S_{2,1}^{(3)}(u) + 2\,S_{2,2}^{(3)}(u)
           \biggr)\,
       \biggr]\,
     \dd u,
\end{align*}
where
\begin{align*}
  S_{0,0}^{(3)}(u)
  &= \sum_{i=0}^\infty
       \frac{u^i}{i!\,J_i^{(3,0)}(u)},
  \\
  S_{1,1}^{(3)}(u)
  &= 3\sum_{i=0}^\infty
       \frac{u^i\,J_i^{(3,1)}(u)}
            {i!\,\bigl(J_i^{(3,0)}(u)\bigr)^2},
  \\
  S_{2,1}^{(3)}(u)
  &= 6\sum_{i=0}^\infty
       \frac{u^i\,J_i^{(3,2)}(u)}
            {i!\,\bigl(J_i^{(3,0)}(u)\bigr)^2},
  \\
  S_{2,2}^{(3)}(u)
  &= \frac{9}{2}
     \sum_{i=0}^\infty
       \frac{u^i\,\bigl(J_i^{(3,1)}(u)\bigr)^2}
            {i!\,\bigl(J_i^{(3,0)}(u)\bigr)^3},
  \\
  T_{1,1}^{(3)}(u)
  &= 3\sum_{i=0}^\infty
        \frac{J_i^{(3,1)}(u)}
             {J_i^{(3,0)}(u)},
  \\
  T_{2,1}^{(3)}(u)
  &= 6\sum_{i=0}^\infty
        \frac{J_i^{(3,2)}(u)}
             {J_i^{(3,0)}(u)},
  \\
  T_{2,2}^{(3)}(u)
  &= \frac{9}{2}
     \sum_{i=0}^\infty
       \frac{\bigl(J_i^{(3,1)}(u)\bigr)^2}
            {\bigl(J_i^{(3,0)}(u)\bigr)^2}.
\end{align*}
\end{example}

The computation results of these examples are found in
Section~\ref{sec:experiments}.

\section{Asymptotic analysis of extreme cases}\label{sec:asympt}

The form of the downlink coverage probability obtained in
Theorem~\ref{thm1} seems complicated and it is hard to derive any
qualitative property from it.
Thus, in this section, we investigate the asymptotic property as
$\theta\to\infty$ for two extreme cases; that is, $m=1$ and
$m\to\infty$ of the Nakagami-$m$ fading.
The case of $m=1$ reduces to the Rayleigh fading and the asymptotic
property is obtained in~\cite{MiyoShir14a} as follows.

\begin{proposition}[Theorem~2 in~\cite{MiyoShir14a}]\label{prp:AsymRayleigh}
The downlink coverage probability for $m=1$ (Rayleigh fading)
satisfies
\begin{equation}\label{eq:prp2-1}
 \lim_{\theta\to\infty}\theta^{1/\beta}\,\Prb(\SIR_o^{(1)}>\theta)
  = \int_0^\infty
      \prod_{j=2}^\infty
        \Exp\Bigl[
          \Bigl(1 + \Bigl(\frac{v}{Y_j}\Big)^\beta\Bigr)^{-1}
        \Bigr]\,
    \dd v,
\end{equation}
where $Y_j\sim\Gam(j, 1)$, $j=1,2,\ldots$, are mutually independent.
\end{proposition}

Since $\Gam(m, 1/m)$ converges weakly to the Dirac measure~$\delta_1$
with mass at the unity as $m\to\infty$, considering the case of
$m\to\infty$ corresponds to ignoring the fading effects.
In this case, we first show that the downlink coverage probability has
the same asymptotic decay rate as the case of $m=1$ and then derive
the relation among the two asymptotic constants.

\begin{theorem}\label{thm:AsymInfty}
In the extreme case of $m\to\infty$, the downlink coverage probability
satisfies
\begin{equation}\label{eq:thm2-1}
  \lim_{\theta\to\infty}\theta^{1/\beta}\,\Prb(\SIR_o^{(\infty)}>\theta)
  = \Exp\Bigl[
       \Bigl(
         \sum_{j=2}^\infty\frac{1}{{Y_j}^\beta}
       \Bigr)^{-1/\beta}
     \Bigr],
\end{equation}
where $Y_j\sim\Gam(j, 1)$, $j=1,2,\ldots$, are mutually independent.

Furthermore, let $c^{(1)}$ and $c^{(\infty)}$ denote the right-hand
sides of \eqref{eq:prp2-1} and \eqref{eq:thm2-1}, respectively.
Then, these asymptotic constants satisfy
\begin{equation}\label{eq:thm2-2}
  c^{(1)} \ge \Gamma\Bigl(1+\frac{1}{\beta}\Bigr)\,c^{(\infty)},
\end{equation}
where $\Gamma$ denotes Euler's gamma function.
\end{theorem}

\begin{IEEEproof}
Applying $H_i=1$ and $\|X_i\|^2\sim Y_i$ for $i\in\N$, and also
$\ell(r)=r^{-2\,\beta}$ in \eqref{eq:SIR} and \eqref{eq:interference},
we have
\begin{align}\label{eq:thm2pr1}
  &\Prb(\SIR_o^{(\infty)} > \theta)
   = \sum_{i=1}^\infty
       \Prb(\SIR_o^{(\infty)} > \theta, \: B_o=i)
  \nonumber\\
  &= \sum_{i=1}^\infty
       \Prb\Bigl(
         \theta\!\sum_{j\in\N\setminus\{i\}}
                    \Bigl(\frac{Y_i}{Y_j}\Bigr)^\beta < 1,\:
         Y_j\ge Y_i, j\in\N
       \Bigr).
\end{align}
First, we consider the summand for $i=1$ on the right-hand side of
\eqref{eq:thm2pr1}.
Applying the density function of $Y_1\sim\mathrm{Exp}(1)$ yields
\begin{align*}
  &\Prb\Bigl(
     \theta\sum_{j=2}^\infty
                \Bigl(\frac{Y_1}{Y_j}\Bigr)^\beta < 1,\:
     Y_j\ge Y_1, j\in\N
   \Bigr)
  \\
  &= \int_0^\infty
       \ee^{-u}\,\,
       \Prb\Bigl(
         \Bigl(\sum_{j=2}^\infty \frac{1}{{Y_j}^\beta}
         \Bigr)^{-1}
         > \theta\,u^\beta,\:
  \\
  &\quad
         Y_j\ge u, j=2, 3, \ldots
       \Bigr)\,
     \dd u
  \\
  &= \theta^{-1/\beta}
     \int_0^\infty
       \ee^{-\theta^{-1/\beta}\,v}\,
       \Prb\Bigl(
         \Bigl(
           \sum_{j=2}^\infty \frac{1}{{Y_j}^\beta}
         \Bigr)^{-1/\beta}
         > v,\:
  \\
  &\quad
         Y_j\ge \theta^{-1/\beta}\,v, j=2, 3, \ldots
       \Bigr)\,
     \dd v,
\end{align*}
where the last equality follows from the substitution of $v =
\theta^{1/\beta}\,u$.
Therefore, since $\ee^{-\theta^{-1/\beta}\,v}\uparrow1$ and
$\ind{\{Y_j\ge\theta^{-1/\beta}\,v\}}\uparrow 1$ a.s.\ as
$\theta\to\infty$, the monotone convergence theorem implies
\begin{align*}
  &\lim_{\theta\to\infty}
     \theta^{1/\beta}\,
     \Prb\Bigl(
       \theta\sum_{j=2}^\infty
                 \Bigl(\frac{Y_1}{Y_j}\Bigr)^\beta < 1,\:
       Y_j\ge Y_1, j\in\N
     \Bigr)
  \\
  &= \int_0^\infty
       \Prb\Bigl(
         \Bigl(
           \sum_{j=2}^\infty\frac{1}{{Y_j}^\beta}
         \Bigr)^{-1/\beta}
         > v
       \Bigr)\,
     \dd v,
\end{align*}
and we obtain the right-hand side of \eqref{eq:thm2-1}.

It remains to show that the summation over $i=2$, $3$, \ldots on the
right-hand side of \eqref{eq:thm2pr1} is $o(\theta^{-1/\beta})$ as
$\theta\to\infty$.
For $i\ge2$, applying the density function of $Y_i\sim\Gam(i, 1)$
yields
\begin{align}\label{eq:thm2pr2}
  &\Prb\Bigl(
     \theta\!\sum_{j\in\N\setminus\{i\}}
                \Bigl(\frac{Y_i}{Y_j}\Bigr)^\beta < 1,\:
     Y_j\ge Y_i, j\in\N
   \Bigr)
  \nonumber\\
  &\le \int_0^\infty
         \frac{\ee^{-u}\,u^{i-1}}{(i-1)!}\,
       \Prb\Bigl(
         \Bigl(
           \sum_{j\in\N\setminus\{i\}}
             \frac{1}{{Y_j}^\beta}
         \Bigr)^{-1}
         > \theta\,u^\beta
       \Bigr)\,
     \dd u
  \nonumber\\
  &= \Exp\Bigl(
       \int_0^{(\theta\,Z_i)^{-1/\beta}}
         \frac{\ee^{-u}\,u^{i-1}}{(i-1)!}\,
       \dd u
     \Bigr),
\end{align}
where
$Z_i= \sum_{j\in\N\setminus\{i\}}(1/{Y_j}^\beta)$.
Therefore, since $\ee^{-u}\le1$,
\begin{equation}\label{eq:thm2pr3}
  \Exp\Bigl(
     \int_0^{(\theta\,Z_i)^{-1/\beta}}
       \frac{\ee^{-u}\,u^{i-1}}{(i-1)!}\,
     \dd u
   \Bigr)
  \le \frac{\theta^{-i/\beta}}{i!}\,\Exp({Z_i}^{-i/\beta}).
\end{equation}
Here, noting that $i\ne1$ and $Y_1\sim\mathrm{Exp}(1)$,
\begin{equation}\label{eq:thm2pr4}
  \Exp({Z_i}^{-i/\beta})
  = \Exp\Bigl[
      \Bigl(
        \sum_{j\in\N\setminus\{i\}} \frac{1}{{Y_j}^\beta}
      \bigr)^{-i/\beta}
    \Bigr]
  \le \Exp\bigl({Y_1}^i\bigr)
  = i!.
\end{equation}
Hence, applying \eqref{eq:thm2pr3} and \eqref{eq:thm2pr4} to
\eqref{eq:thm2pr2}, we obtain
\begin{align*}
  &\theta^{1/\beta}
   \sum_{i=2}^\infty
     \Prb\Bigl(
       \theta\!
       \sum_{j\in\N\setminus\{i\}}
         \Bigl(\frac{Y_i}{Y_j}\Bigr)^\beta < 1,\:
       Y_j\ge Y_i, j\in\N
     \Bigr)
  \\
  &\le \theta^{1/\beta}
       \sum_{i=2}^\infty \theta^{-i/\beta}
   =   \frac{1}{\theta^{1/\beta}-1}
   \to0\quad\text{as $\theta\to\infty$,}
\end{align*}
which completes the proof of the first part in Theorem~\ref{thm:AsymInfty}.

For the second part of the theorem, since $Y_j$, $j\in\N$, are
mutually independent, the right-hand side of \eqref{eq:prp2-1} reduces
to
\begin{align*}
  c^{(1)}
  &= \int_0^\infty
       \Exp\biggl(
         \exp\biggl\{
           -\sum_{j=2}^\infty
              \log\Bigl(
                1 + \Bigl(\frac{v}{Y_j}\Bigr)^\beta
              \Bigr)
         \biggr\}
       \biggr)\,
     \dd v
  \\
  &\ge \int_0^\infty
         \Exp\bigl( \ee^{-v^\beta\,Z_1} \bigr)\,
       \dd v,
\end{align*}
where $Z_1=\sum_{j=2}^\infty(1/{Y_j}^\beta)$ and the inequality follows
from $\log(1+x)\le x$.
Thus, substituting $t=v^\beta\,Z_1$,
\begin{align*}
  \int_0^\infty
    \Exp\bigl( \ee^{-v^\beta\,Z_1} \bigr)\,
  \dd v
  &= \frac{1}{\beta}\int_0^\infty \ee^{-t}\,t^{1/\beta-1}\,\dd t\,
     \Exp\bigl({Z_1}^{-1/\beta}\bigr)
  \\
  &= \Gamma\Bigl(1+\frac{1}{\beta}\Bigr)\,c^{(\infty)}.
\end{align*}
\end{IEEEproof}

Note in \eqref{eq:thm2-2} that $\Gamma(1+1/\beta)\uparrow1$ as
$\beta\to1$ and $\beta\to\infty$, and it takes the minimum value of
the gamma function on the positive domain
$\Gamma(\gamma_{\min})=0.8856031944\cdots$ at
$1+1/\beta=\gamma_{\min}=1.4616321449\cdots$; that is, the coefficient
on the right-hand side of \eqref{eq:thm2-2} is somewhat close to 1.

\section{Numerical experiments}\label{sec:experiments}

\begin{figure}%
\begin{center}
\includegraphics[width=\linewidth]{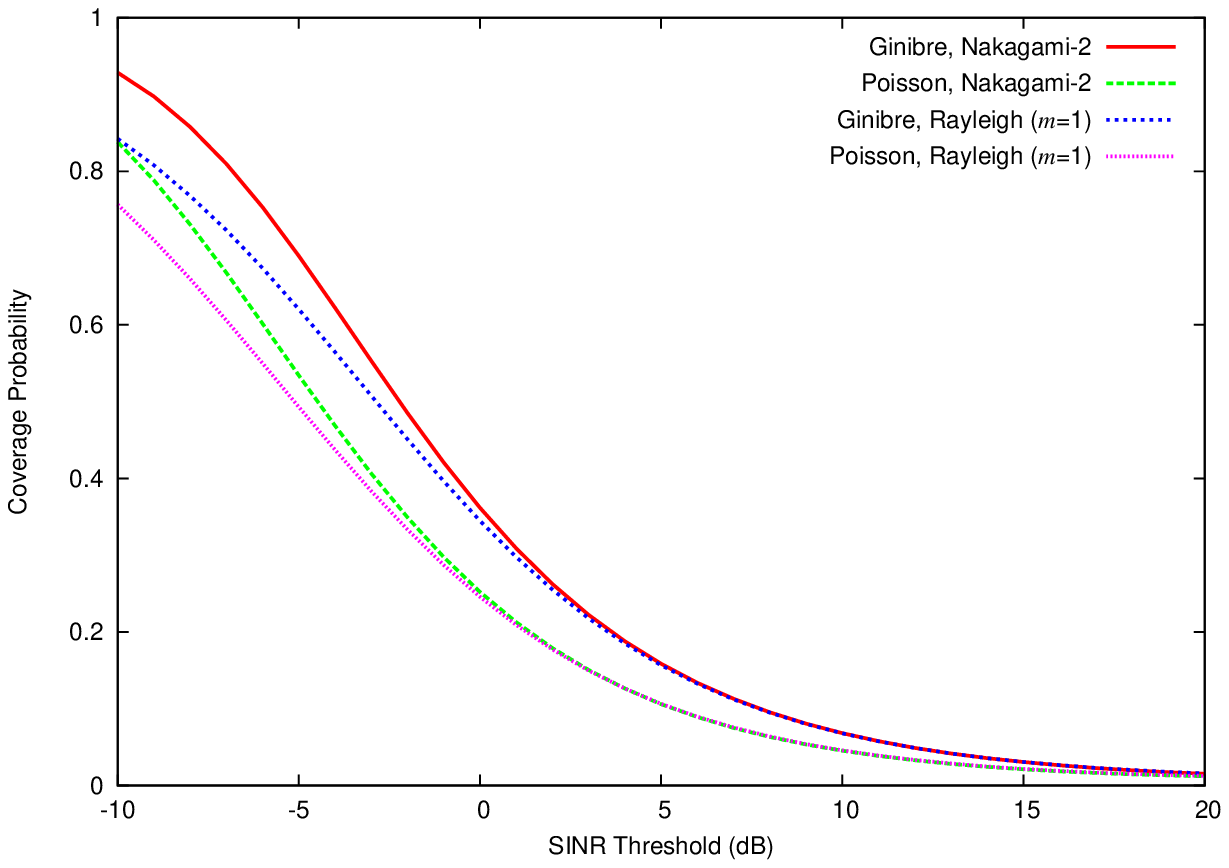}
\\[1ex]
\includegraphics[width=\linewidth]{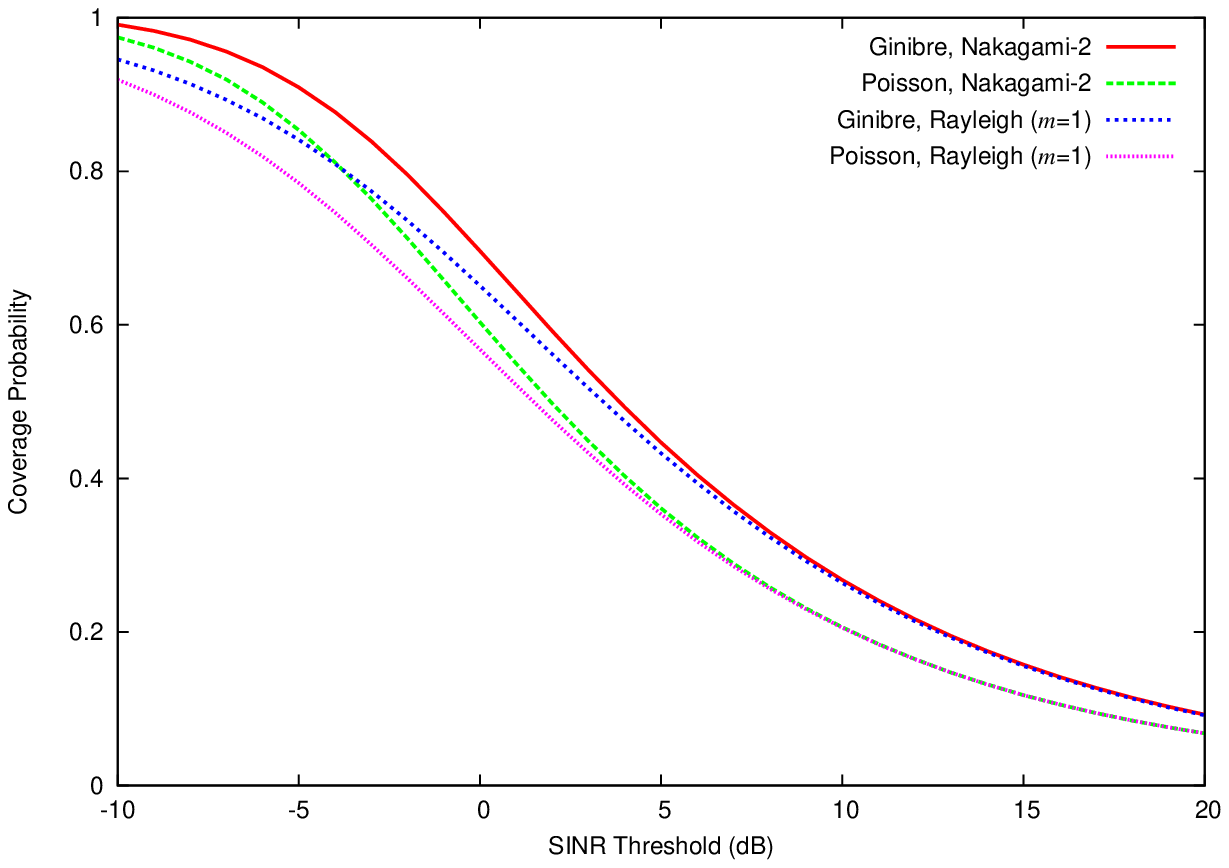}
\end{center}
\caption{Comparison of downlink coverage probability between the GPP
  and PPP based models for $\beta=1.25$ (top) and $\beta=2.0$
  (bottom).}\label{fig:compare_withP}
\end{figure}

The results of some numerical experiments are presented.
The first experiment is the comparison between the GPP and PPP based
models.
In Figure~\ref{fig:compare_withP}, the downlink coverage probability
over Nakagami-$2$ fading is plotted for both the GPP and PPP based
models, where two cases $\beta=1.25$ and $\beta=2.0$ (that is,
$\ell(r)=r^{-2.5}$ and $\ell(r)=r^{-4}$) are computed.
For comparison, the corresponding results over the Rayleigh fading,
which are obtained in \cite{MiyoShir14a}, are also displayed there.
As in the Rayleigh fading case, the downlink coverage probability for
the GPP based model dominates that for the PPP based model in the
Nakagami-$m$ fading case.
Furthermore, for both the GPP and PPP based models, the coverage
probability seems asymptotically invariant in the value of $m$.
This observation is, however, doubtful from the results of
Lemma~\ref{lem:icx} and Theorem~\ref{thm:AsymInfty}.
Thus, this is further investigated in the next experiment.

\begin{figure}%
\begin{center}
\includegraphics[width=\linewidth]{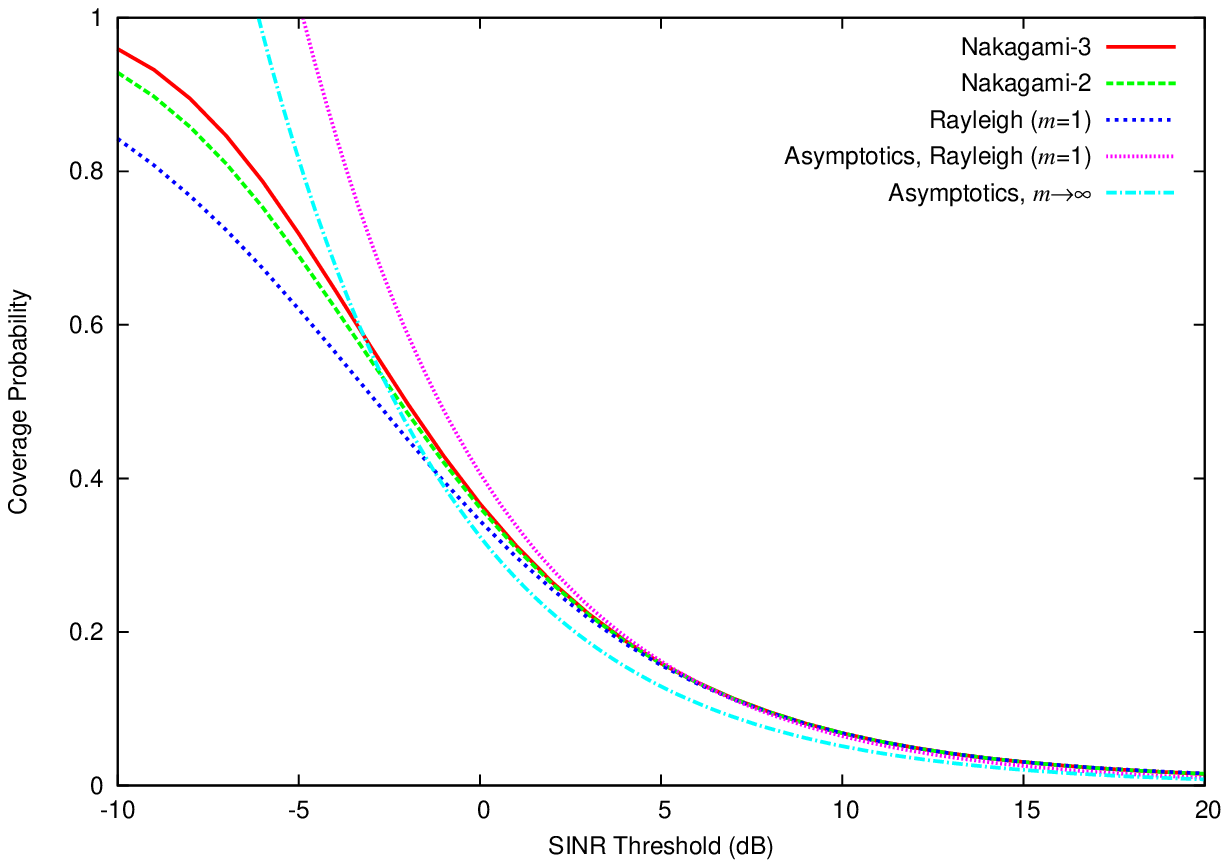}
\\[1ex]
\includegraphics[width=\linewidth]{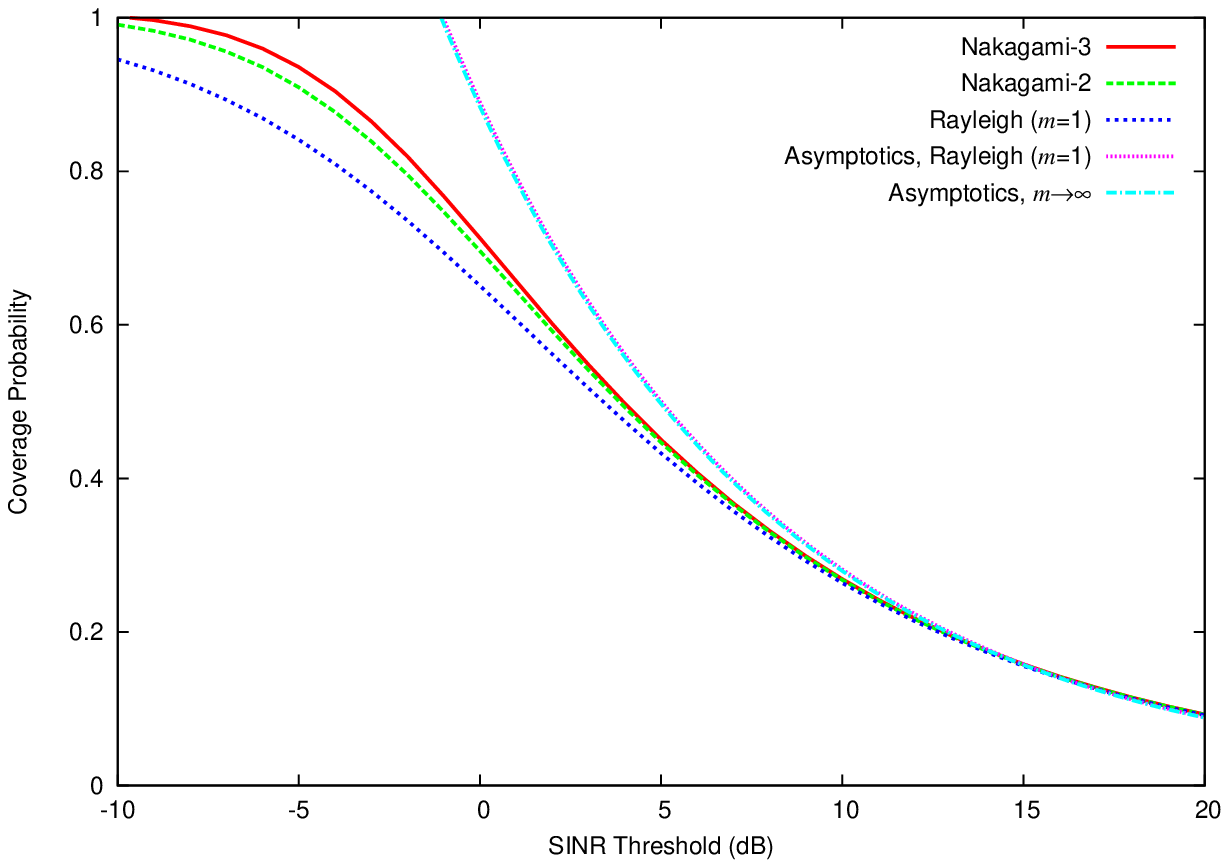}
\end{center}
\caption{Comparison of downlink coverage probability in terms of the
  value of $m$ for $\beta=1.25$ (top) and $\beta=2.0$
  (bottom).}\label{fig:compare_m}
\end{figure}

In the second experiment, we compare the downlink coverage probability
in terms of $m$ of the Nakagami-$m$ fading.
The coverage probability for $m=1$, $2$ and $3$ is plotted in
Figure~\ref{fig:compare_m}.
The asymptotic results obtained in Proposition~\ref{prp:AsymRayleigh}
and Theorem~\ref{thm:AsymInfty} are also drawn in the same figure,
where the asymptotic constant~$c^{(\infty)}$ for the case of
$m\to\infty$ is estimated from 100 independent samples of
$\bigl(\sum_{j=2}^\infty 1/{Y_j}^\beta\bigr)^{-1/\beta}$ in
\eqref{eq:thm2-1}.
From the figure, indeed the downlink coverage probability seems
increasing in $m=1$, $2$ and $3$.
However, as $m\to\infty$, the asymptotic tail of the coverage
probability is smaller than the others particularly for the value of
$\beta$ close to $1$.
This observation agrees with the results of Lemma~\ref{lem:1} and
Theorem~\ref{thm:AsymInfty}.
Anyway, a further investigation would be required concerning the
impact of the value of $m$ of the Nakagami-$m$ fading.

\section*{Acknowledgment}

The first author's work was supported in part by Japan Society for the
Promotion of Science (JSPS) Grant-in-Aid for Scientific Research (C)
25330023.
The second author's work was supported in part by JSPS Grant-in-Aid
for Scientific Research (B) 26287019.



\begin{thebibliography}{10}
\providecommand{\url}[1]{#1}
\csname url@samestyle\endcsname
\providecommand{\newblock}{\relax}
\providecommand{\bibinfo}[2]{#2}
\providecommand{\BIBentrySTDinterwordspacing}{\spaceskip=0pt\relax}
\providecommand{\BIBentryALTinterwordstretchfactor}{4}
\providecommand{\BIBentryALTinterwordspacing}{\spaceskip=\fontdimen2\font plus
\BIBentryALTinterwordstretchfactor\fontdimen3\font minus
  \fontdimen4\font\relax}
\providecommand{\BIBforeignlanguage}[2]{{%
\expandafter\ifx\csname l@#1\endcsname\relax
\typeout{** WARNING: IEEEtran.bst: No hyphenation pattern has been}%
\typeout{** loaded for the language `#1'. Using the pattern for}%
\typeout{** the default language instead.}%
\else
\language=\csname l@#1\endcsname
\fi
#2}}
\providecommand{\BIBdecl}{\relax}
\BIBdecl

\bibitem{BaccBlas09a}
F.~Baccelli and B.~B{\l}aszczyszyn, ``Stochastic geometry and wireless
  networks, {Volume~I:\ Theory},'' \emph{Foundations and Trends~(R) in
  Networking}, vol.~3, pp. 249--449, 2009.

\bibitem{BaccBlas09b}
------, ``Stochastic geometry and wireless networks, {Volume~II:\
  Applications},'' \emph{Foundations and Trends~(R) in Networking}, vol.~4, pp.
  1--312, 2009.

\bibitem{Haen13}
M.~Haenggi, \emph{Stochastic Geometry for Wireless networks}.\hskip 1em plus
  0.5em minus 0.4em\relax Cambridge University Press, 2013.

\bibitem{Mukh14}
S.~Mukherjee, \emph{Analytical Modeling of Heterogeneous Cellular Networks:
  Geometry, Coverage, and Capacity}.\hskip 1em plus 0.5em minus 0.4em\relax
  Cambridge University Press, 2014.

\bibitem{MiyoShir14a}
N.~Miyoshi and T.~Shirai, ``A cellular network model with {Ginibre} configured
  base stations,'' \emph{Advances in Applied Probability}, vol.~46, pp.
  832--845, 2014.

\bibitem{TorrLeon14}
G.~L. Torrisi and E.~Leonardi, ``Large deviations of the interference in the
  {Ginibre} network model,'' \emph{Stochastic Systems}, vol.~4, pp. 1--33,
  2014.

\bibitem{NakaMiyo14}
I.~Nakata and N.~Miyoshi, ``Spatial stochastic models for analysis of
  heterogeneous cellular networks with repulsively deployed base stations,''
  \emph{Performance Evaluation}, vol.~78, pp. 7--17, 2014.

\bibitem{MiyoShir14b}
N.~Miyoshi and T.~Shirai, ``Cellular networks with {$\alpha$-Ginibre}
  configurated base stations,'' in \emph{The Impact of Applications on
  Mathematics: Proceedings of the Forum of Mathematics for Industry
  2013}.\hskip 1em plus 0.5em minus 0.4em\relax Springer, 2014, pp. 211--226.

\bibitem{KobaMiyo14}
T.~Kobayashi and N.~Miyoshi, ``Uplink cellular network models with {Ginibre}
  deployed base stations,'' in \emph{26th International Teletraffic Congress
  (ITC)}, 2014.

\bibitem{LiBaccDhilAndr14}
Y.~Li, F.~Baccelli, H.~S. Dhillon, and J.~G. Andrews, ``Statistical modeling
  and probabilistic analysis of cellular networks with determinantal point
  processes,'' 2014, arXiv:1412.2087~[cs.IT].

\bibitem{Sosh00}
A.~Soshnikov, ``Determinantal random point fields,'' \emph{Russian Mathematical
  Surveys}, vol.~55, pp. 923--975, 2000.

\bibitem{ShirTaka03}
T.~Shirai and Y.~Takahashi, ``Random point fields associated with certain
  {Fredholm} {determinants~I:\ Fermion, Poisson and Boson processes},''
  \emph{Journal of Functional Analysis}, vol. 205, pp. 414--463, 2003.

\bibitem{HougKrisPereVira09}
J.~B. Hough, M.~Krishnapur, Y.~Peres, and B.~Vir\'ag, \emph{Zeros of Gaussian
  Analytic Functions and Determinantal Point Processes}.\hskip 1em plus 0.5em
  minus 0.4em\relax American Mathematical Society, 2009.

\bibitem{Mukh11}
S.~Mukherjee, ``Downlink {SINR} distribution in a heterogeneous cellular
  wireless network with {max-SINR} connectivity,'' in \emph{49th Annual
  Allerton Conference on Communication, Control and Computing}, 2011, pp.
  1649--1656.

\bibitem{MadhRestLiuBrow12}
P.~Madhusudhanan, J.~G. Restrepo, Y.~Liu, and T.~X. Brown, ``Downlink coverage
  analysis in a heterogeneous cellular network,'' in \emph{2012 IEEE Global
  Communications Conference (GLOBECOM)}, 2012, pp. 4170--4175.

\bibitem{KeelBlasKarr13}
H.~Keeler, B.~B{\l}aszczyszyn, and M.~Karray, ``{SINR-based} $k$-coverage
  probability in cellular networks with arbitrary shadowing,'' in \emph{2013
  IEEE International Symposium on Information Theory Proceedings (ISIT)}, 2013,
  pp. 1167--1171.

\bibitem{DiReGuidCora13}
M.~D. Renzo, A.~Guidotti, and G.~E. Corazza, ``Average rate of downlink
  heterogeneous cellular networks over generalized fading channels:\ {A}
  stochastic geometry approach,'' \emph{IEEE Transactions on Communications},
  vol.~61, pp. 3050--3071, 2013.

\bibitem{Naka60}
M.~Nakagami, ``The $m$-distribution---{A} general formula of intensity
  distribution of rapid fading,'' in \emph{Statistical Methods in Radio Wave
  Propagation: Proceedings of a Symposium Held at the University of California,
  Los Angeles, June 18--20, 1958}, W.~C. Hoffman, Ed.\hskip 1em plus 0.5em
  minus 0.4em\relax Pergamon Press, 1960, pp. 3--36.

\bibitem{HuanPapaVenk12}
H.~Huang, C.~B. Papadias, and S.~Venkatesan, \emph{MIMO Communication for
  Cellular Networks}.\hskip 1em plus 0.5em minus 0.4em\relax Springer, 2012.

\bibitem{DhilKounAndr13}
H.~S. Dhillon, M.~Kountouris, and J.~G. Andrews, ``Downlink {MIMO HetNets:
  Modeling}, ordering results and performance analysis,'' \emph{IEEE
  Transactions on Wireless Communications}, vol.~12, pp. 5208--5222, 2013.

\bibitem{GuptDhilVishAndr14}
A.~K. Gupta, H.~S. Dhillon, S.~Vishwanath, and J.~G. Andrews, ``Downlink
  coverage probability in {MIMO HetNets} with flexible cell selection,'' in
  \emph{2014 IEEE Global Communications Conference (GLOBECOM)}, 2014, pp.
  1534--1539.

\bibitem{AndrBaccGant11}
J.~G. Andrews, F.~Baccelli, and R.~K. Ganti, ``A tractable approach to coverage
  and rate in cellular networks,'' \emph{IEEE Transactions on Communications},
  vol.~59, pp. 3122--3134, 2011.

\bibitem{MullStoy02}
A.~{M\"{u}ller} and D.~Stoyan, \emph{Comparison Methods for Stochastic Models
  and Risks}.\hskip 1em plus 0.5em minus 0.4em\relax John Wiley \& Sons, 2002.

\bibitem{Kost92}
E.~Kostlan, ``On the spectra of {Gaussian} matrices,'' \emph{Linear Algebra and
  its Applications}, vol. 162--164, pp. 385--388, 1992.

\bibitem{Olve00}
P.~J. Olver, \emph{Applications of Lie Groups to Differential Equations}.\hskip
  1em plus 0.5em minus 0.4em\relax Springer, 2000.

\bibitem{Roma80}
S.~Roman, ``The formula of {Fa\`a di Bruno},'' \emph{American Mathematical
  Monthly}, vol.~87, pp. 805--809, 1980.

\end{thebibliography}

\end{document}